Astro2020 APC White Paper
**Findings and Recommendations from the American Astronomical Society (AAS) Committee on the Status of Women in Astronomy: Towards Eliminating Harassment in Astronomy**
Thematic Area: State of the Profession Considerations

Principal Author:

Nicolle Zellner, Albion College, nzellner@albion.edu

Co-Authors:

JoEllen McBride, University of Pennsylvania, joellen.mcbride@gmail.com
Nancy Morrison, University of Toledo, nancyastro126@gmail.com
Alice Olmstead, Texas State University, alice.olmstead@txstate.edu
Maria Patterson, High Alpha, maria.t.patterson@gmail.com
Gregory Rudnick, University of Kansas, grudnick@ku.edu
Aparna Venkatesan, University of San Francisco, avenkatesan@usfca.edu
Heather Flewelling, University of Hawaii Institute for Astronomy, heather@ifa.hawaii.edu
David Grinspoon, Planetary Science Institute, grinspoon@psi.edu
Jessica D. Mink, Harvard University Center for Astrophysics, jmink@cfa.harvard.edu
Christina Richey, Jet Propulsion Laboratory, christina.r.richey@jpl.nasa.gov
Angela Speck, University of Missouri, speckan@missouri.edu
Cristina A. Thomas, Northern Arizona University, Cristina.Thomas@nau.edu
Sarah E. Tuttle, University of Washington, tuttlese@uw.edu

Endorsers:
Adam Burgasser, UC San Diego, aburgasser@ucsd.edu
Kim Coble, San Francisco State University, kcoble@sfsu.edu
Courtney Dressing, University of California, Berkeley, dressing@berkeley.edu
Emily Martin, UC Santa Cruz, emilymartin@ucsc.edu
Jane Rigby, NASA GSFC, Jane.Rigby@nasa.gov
Lia Corrales, University of Michigan, liac@umich.edu
Shawn D. Domagal-Goldman, NASA GSFC, shawn.goldman@nasa.gov
Stephen Lawrence, Hofstra University, Stephen.Lawrence@hofstra.edu

**Executive Summary**

The Committee on the Status of Women in Astronomy (CSWA) is calling on federal science funding agencies, in their role as the largest sources of funding for astronomy in the United States, to take actions that will end harassment, particularly sexual harassment, in astronomical workplaces. Funding agencies can and should lead the charge to end harassment in astronomy by the 2030 Astrophysics Decadal Survey. Anecdotal and quantitative evidence, gathered both by the CSWA and other groups, shows that harassment is prevalent and damaging for women and minority astronomers and those in related fields. Actions recommended herein will increase the rate of reporting of harassment to agencies and improve their ability to investigate and take action against harassers. We also recommend that agencies participate in harassment prevention by creating and implementing the best anti-harassment education possible. Key recommendations are:

- Federal agencies should improve their ethics policies by making harassment a form of scientific misconduct.
- Federal agencies should mandate that institutions report to them when a funded Principal Investigator (PI) or co-Principal Investigator (co-PI) is found to be a perpetrator of harassment.
- Federal funding agencies should provide online guides to help scientists identify harassment and connect them to the right resources for making confidential or official reports.
- Federal agencies should create and ensure the implementation of anti-harassment trainings by making them a requirement of receiving grant funding.

1. **Findings and Overview**

1.1 Purpose
*This paper proposes actionable, evidence-supported policies that will help end harassment.*

This paper will focus on ending harassment, while a sister paper focuses on career development issues. We acknowledge that these areas are connected, and addressing each will have profound impacts on the other. The contents of Sections 1.1 and 1.2 and the first paragraph of Section 1.3 are repeated in both papers.

The CSWA was created in 1979 and was charged with making practical recommendations to the AAS council on what can be done to improve the status of women in astronomy. The CSWA's scope has expanded to include all bodies that influence the work lives of astronomers, including research facilities, academic institutions, the federal government, and others[1]. The CSWA is submitting two reports on the pressing issues affecting women and minority astronomers, and recommending policy changes that fit the needs of our community in response to the decadal survey's call for white papers on the state of the astronomy profession.

1.2 Methodology

In Spring 2019, the CSWA conducted a survey to assess astronomers' perspectives on policies in four areas of concern: harassment and bullying, creating inclusive environments, professional development, and ethics. The survey included 53 Likert-scale questions that allowed astronomers to rate the effectiveness of policy actions that could be undertaken by relevant stakeholders, and 17 free response opportunities that they used to explain their answers. We received over 340 responses to the survey. No personally identifiable information was collected. Although we acknowledge the disadvantage of being unable to categorize our respondents' opinions by demographics, we believe the anonymous nature of the survey increased astronomers' willingness to take the survey and write candid free responses. Relevant expertise and data in the field of STEM equity also inform our recommendations.

1.3 Progress for Women and Minorities in Astronomy is Slowing
*The number of women earning astronomy PhDs is increasing, but at a decreasing rate.*

According to the National Science Foundation (NSF), the number of women earning physics and astronomy doctoral degrees is increasing, but at a decreasing rate. The number of women who received doctoral degrees in physics and astronomy increased by 21.3% from 2008 to 2012, but by only 8.1% from 2013 to 2017[2]. Additionally, the American Institute of Physics (AIP) reports consistent decreases in the number of women earning astronomy bachelor's degrees over the past ten years[3]. These trends are consistent with Figure 1, from Ivie & Porter[3].

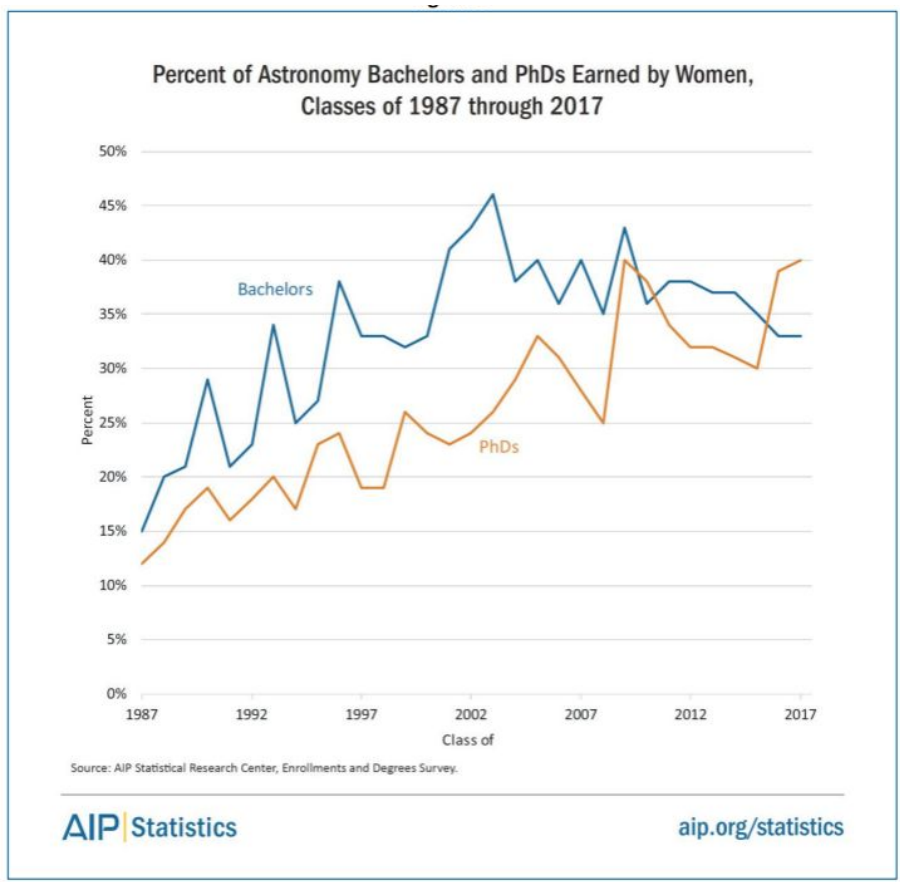

**Figure 1.** The number of women earning physics and astronomy PhDs is increasing at a decreasing rate, and the number of women earning astronomy bachelor's degrees is decreasing. These data indicate there are significant obstacles on the path towards equal representation of all genders.

1.4 Harassment Persists for Women and Underrepresented Minorities (URMs) in Astronomy
*The literature shows that harassment in astronomy has a significant impact on women, and an even higher impact on URM women.*

Harassment persists as a problem throughout STEM and academia. The landmark 2018 report on sexual harassment in STEM fields from the National Academies of Science, Engineering, and Medicine[4] found that only 6% of women graduate students and faculty members who are targets of sexual harassment choose to report the incidents to their institutions, despite the fact that around *50% of all women graduate students and faculty members have experienced some form of sexual harassment.*

Harassment permeates the experiences of women and minority astronomers. In their 2017 study using results from the Longitudinal Survey of Astronomy Graduate Students, Ivie, White, and Chu[5] found that, compared to non-URM men, non-URM women are 8 times more likely, and URM women are 20 times more likely, to experience discrimination or harassment. In another survey of 474 astronomers and planetary scientists, 40% of women of color reported feeling unsafe in the workplace as a result of their gender or sex, and 28% of women of color reported feeling unsafe as a result of their race. 18% of women of color and 12% of white women reported skipping professional events because they did not feel safe attending, identifying a significant loss of career opportunities due to a hostile climate[6].

Of respondents to the American Physical Society's 2016 LGBT Climate Survey, about 15% of LGBT men, 25% of LGBT women, and 30% of gender-nonconforming individuals characterized the overall climate of their department or division as "uncomfortable" or "very uncomfortable."[7] The urgency of the situation is summed up well by the words of one of our survey respondents, who wrote, "Ignoring intersectionality should not be treated as an option." Taking an intersectional approach to anti-harassment education is essential.

Even as more women enter the field, they are underrepresented in key astronomical workplaces. Trends reported by the AIP show no growth in the number of African American women faculty in astronomy in the United States from 2008-2016[3]. According to a study of NASA mission teams, from 2001 through 2016, the percentage of women involved in spacecraft science teams remained flat, at 15.8%[8]. Ivie, White, and Chu[5] report that women and URMs are more likely than men to choose to work outside of astronomy. Our survey respondents reported that harassment is one of the most significant factors preventing women and minorities from continuing in the field and attaining high ranks.

The NASEM report defines three categories of sexual harassment: sexual coercion, unwanted sexual attention, and gender harassment. Their findings indicate that the vast majority of sexual harassment is gender harassment, which they define as verbal and nonverbal behaviors that convey hostility, objectification, exclusion, or second-class status towards members of one's gender. It is imperative to take action to decrease not only the first two types of sexual harassment, but also gender harassment, which can cause untenable exhaustion and stress for targets of harassment and bystanders of all genders in any work environment. An environment characterized by the toxic stress caused by widespread gender harassment drives people out of the workplace, and in many cases, out of their field forever[4].

There is also a pressing need to address implicit bias. Of our survey respondents, 76% agreed that encouraging institutions to implement ways to mitigate implicit bias would be a "somewhat effective" or "very effective" policy in improving their workplaces. Implicit bias has a marked impact on the field. A study by Caplar et al.[9] shows that astronomy papers authored by women receive 10% fewer citations than would be expected if the papers were written by men. This sentiment is echoed in a current bill, H.R. 2528, The STEM Opportunities Act of 2019: "Decades of cognitive psychology research reveal that most people carry prejudices of which they are unaware but that nonetheless play a large role in evaluations of people and their work. Unintentional biases and outmoded institutional structures are hindering the access and advancement of women, minorities, and other groups historically underrepresented in STEM."[10]

1.5 Implement Policy Action to End Harassment
*Federal agencies can and should take action to end harassment.*

This June, the House of Representatives Committee on Space, Science, and Technology held a hearing to investigate efforts to combat sexual harassment in STEM fields. In her opening statement, Chairwoman Eddie Bernice Johnson said, "The public investment in research needs to draw on all of our nation's talent to return the best possible science for the benefit of society. To reach this goal, we must do more to ensure that all researchers have access to a safe work environment."[11] Federal agencies are a powerful stakeholder in the scientific enterprise, and they must take action to end harassment.

## 2. Ethics
*Strengthening ethics policies is an essential step to end harassment in STEM.*

2.1 Make Harassment a Form of Research Misconduct
*Federal agencies should make harassment a form of research misconduct in order to reform agency priorities and strengthen their response to harassment.*

The CSWA urges agencies to add harassment to their definitions of research misconduct, making it pursuant to the same enforcement processes and penalties. Presently, NASA defines research misconduct as "fabrication, falsification, or plagiarism in proposing, performing, or reviewing research, or in reporting research results," and NSF employs a similar definition. This definition omits many behaviors, including sexual harassment, that have a direct impact on the integrity of research. The NASEM report details the series of events leading to this narrow definition of research misconduct and reveals that there is no justification for the widespread exclusion of harassment from the definition of research misconduct beyond the existence of other laws and policies that prohibit harassment[4].

Of our survey respondents, 82% rated adding harassment to the list of actions that constitute research misconduct as a "somewhat effective" or "very effective" policy to improve professional ethics in the workplace. The American Geophysical Union (AGU) has already made this change[12].

Elevating harassment to the level of research misconduct will increase the resources available for investigating harassment allegations. The Government Accountability Office (GAO) found that five major funding agencies, including NASA and NSF, handle internal and external harassment allegations through their civil rights or diversity offices. Officials in these offices usually handle several initiatives related to diversity and inclusion and are not always trained in harassment investigation or readily available when such issues arise. In contrast, research misconduct investigations are conducted by dedicated staff; for example, at NASA, the Office of the Inspector General investigates research misconduct within NASA funded projects[13].

Bringing harassment under the purview of research misconduct means diversity and inclusion offices *and* investigative offices may have authority over harassment cases. These individuals' and groups' roles and responsibilities will need to be reorganized in order to designate who will investigate harassment and engage those teams in extensive training. This transition will allow agencies to determine the best way to respond to harassment cases given a larger array of internally available resources. We recommend agencies increase their resources for investigating harassment in order to take on as many cases as possible. They should review and strengthen their procedures for deciding which cases to pursue and which allegations to delegate back to the perpetrator's institution or employer. In time, the increased likelihood of agency intervention will incentivize institutions take their internal investigations more seriously.

As of now, most federal policies treat harassment and research misconduct as separate issues, implying that harassment does not impact the quality of science. Federal agencies should lean forward and change the definition of research misconduct, which will push themselves, professional societies, academic research institutions, and other groups to attend to harassment issues with the same degree of attention and resources as other forms of misconduct.

2.2 Implement Mandatory Reporting of Harassment to Agencies
*Implementing mandatory reporting of harassment findings to agencies will increase reporting rates and engage more resources to investigate and sanction harassers.*

The NSF has implemented a new term and condition that mandates NSF-funded institutions report to NSF when a funded Principal Investigator (PI) or co-PI is found guilty of harassment or assault, or when action is taken against them pending an investigation of accusations thereof. The NSF has stated that upon receiving such information, it is in its power "to initiate the substitution or removal of the PI or co-PI, reduce the award funding amount, or where neither of those previous options is available or adequate, to suspend or terminate the award."[14] NASA is poised to implement a similar term and condition by the end of the year[13].

The CSWA supports the implementation of these terms and conditions, and the provisions of the bill H.R. 36, the Combating Sexual Harassment in Science Act of 2019, that relate to such terms and conditions. The proposed law requires the development and implementation of harassment reporting terms and conditions, like the one used by NSF, at all major science funding agencies. It instructs them to share sexual harassment findings among agencies, making it possible for PIs who receive funding from multiple federal sources to be cut off from all sources of funding. The bill encourages agencies to pursue consistency among their policies and procedures regarding harassment through a working group managed by the Office of Science and Technology Policy (OSTP)[15].

A report from the GAO details that NSF has seen an increased number of harassment reports, requests for information on how to report, requests for training, and other inquiries related to preventing and dealing with harassment since the implementation of the new term and condition in October 2018[13]. As more members of the science community have become aware that NSF has equipped itself to take action to remove perpetrators from power, more are showing an interest in reporting. This is an extremely significant finding, since, as stated previously, only 6% of graduate students and faculty members who are targets of sexual harassment choose to report. Several CSWA survey respondents wrote that members of our community have been discouraged from reporting because often, nothing is done about the reports, and in many cases, the perpetrator continues to receive funding. Demonstrated willingness to sanction those with findings against them, along with clear reporting avenues, will increase the volume of reports to funding agencies.

2.3 Create Action Guides
*Federal agencies should create action guides on harassment, which will help reduce the burden on targets of harassment, witnesses, and their allies of navigating the environment of harassment response, and increase reporting rates.*

To support increased reporting and investigation of harassment, federal funding agencies should solicit proposals to create online guides to help scientists identify harassment and connect them to the right resources for making confidential or official reports. Our survey respondents specifically requested guides created by communications professionals and anti-harassment experts that are easy to find and use. A strong guide will help scientists determine the aspects of a possibly toxic work environment that may constitute harassment by providing brief and clear descriptions of the ways harassment manifests, including the three types of sexual harassment described by the NASEM report. It will also help those affected by harassment decide on a course of action by explaining how a report to their institution, professional society, or relevant funding agency is likely to proceed, and the potential disciplinary consequences that may result from an investigation. This guide should be posted via an easy to find link on the homepages of agency websites and advertised to academic institutions and professional societies.

### 3. Changing Workplace Culture Through Trainings

*Federal agencies should facilitate the creation and implementation of well-crafted, engaging, and thorough anti-harassment training.*

3.1 Solicit Proposals to Develop Trainings

*Federal agencies should set the standard for anti-harassment education by creating and implementing the best possible anti-harassment trainings. These trainings should be created by experts, be specific to STEM fields, and emphasize bystander intervention strategies.*

Eradicating harassment and implicit bias necessitates cultural change, and institutions often address this need through trainings that are meant to increase awareness of what harassment looks like, how it can be prevented, and the resources available to targets of harassment. There is no standardization among institutions as to the content, length, and form of trainings[16]. The following recommendations address the need for better and more widespread education on these issues.

The CSWA recommends that federal funding agencies solicit proposals to develop research-supported trainings that can be implemented at institutions, society meetings, and anywhere else scientists gather. Federal agencies should make the trainings they develop available to small and under-resourced groups and institutions at low or no cost.

The NASEM report includes two important findings regarding trainings that federal agencies should keep in mind. First, a poor or inadequate training is worse than no training at all, because it sends the message that the institution does not take harassment and discrimination seriously. Second, trainings should focus on changing behaviors, which has been found to be more effective than changing beliefs. For example, bystander intervention training is highly effective in teaching trainees to recognize and address problematic behavior[4].

In our survey, we asked astronomers to rate the effectiveness of different types of trainings and to describe their experiences with trainings. From these data, we gathered a list of six traits of effective trainings that federal agencies should use when soliciting proposals for anti-harassment and implicit bias trainings.

1. *Bystander intervention should always be taught as a core component of anti-harassment training.* 74% of respondents to our survey said utilizing bystander intervention training would be a "somewhat effective" or "very effective" strategy to reduce harassment.
2. *Trainings should be developed by experts and administered by training professionals.* Our survey respondents emphasized that they could not address harassment without proper professional support, and that while they are able to report on their experiences, they do not have the expertise to create adequate solutions.
3. *Content should be specific to the power structures present in the workplaces of the target audience.* Several of our survey respondents felt that the anti-harassment trainings available to them were not appropriate for physical science workplaces or academic astronomy. Federal science funding agencies should use their resources to gather case studies of how toxic situations manifest in science workplaces and use this information to inform trainings.
4. *Trainings should demonstrate how the possession of more than one marginalized identity may lead to increased experiences of harassment.* Our survey respondents stressed the importance of increasing awareness of intersectional issues.
5. *In-person trainings should center around active engagement. If online trainings are employed, they should present the participant with thought-provoking "gray area" examples.* Ideally, all scientists would have access to in-person trainings, but we understand that this will not always be the case. Experts in online education should be employed to create the most effective remote trainings possible.
6. *Trainings should include a balanced combination of anonymized, real-life examples of harassment and discrimination and reputable data on prevalence and impact.* Scientific audiences are more likely to buy into a narrative supported by both quantitative and qualitative data.

3.3 Make Training and Awareness a Requirement of Funding
*Federal agencies should help ensure that the projects they fund create safe workplaces for women and URMs by making training and awareness a requirement of receiving grant funding.*

Federal funding agencies should require that project proposals include the PI's plan for maintaining a harassment-free environment. An adequate plan should ensure all team members attend, on a yearly basis, one or more trainings that meet at least some of the standards put forth in section 3.2. This may involve using the resources already available at the grantee institution or, if the institution does not provide adequate resources, training from a private source or an agency through the low-to-no cost program proposed in section 3.2. To the extent possible, agencies should coordinate their requirements to accommodate projects that receive funding from multiple sources.

A strong plan will demonstrate the PI's awareness of the resources available to targets of harassment at the federal, field-wide, and institutional levels, and their intention to bring awareness of the proper resources to their team. It should also include the PI's plan to maintain full cognizance of their team's culture and respond if their work environment shows signs of ambient or explicit discrimination or harassment.

Federal funding agencies should encourage institutions and departments to engage with the American Association for the Advancement of Science's (AAAS) SEA Change program by allowing PIs at institutions and departments with an AAAS SEA Change award to present condensed anti-harassment plans. The SEA Change award is a signal that an institution is committed to changing their culture and providing resources to its community, and can be used as an indicator to streamline the proposal process[17].

### 4. Concluding Remarks

*Federal agencies should take action because it is the moral path forward and the path to better science.*

We invite the federal agencies, as the largest sources of funding for U.S. astronomy, to help lead our collective charge towards ending harassment by Astro2030 at the very latest. The case for the prevalence of sexual harassment in STEM fields, and astronomy specifically, is undeniable. When astronomers have no choice but to work in environments characterized by harassment and abuse, they often fail to produce the groundbreaking, awe-inspiring work that pushes astronomy forward. As a community, we cannot afford to let these conditions persist for another decade. We call upon federal agencies to expand the definition of research misconduct to include harassment, to monitor and sanction PIs who are perpetrators of harassment, and to ensure the implementation of thorough anti-harassment education. Equity and inclusion, unlike the funding of projects and missions, is not a zero sum game: improving the diversity of astronomy as a field will lead to high-quality, impactful science. We look forward to the scientific achievements of a workforce that fully utilizes the talents of all its members.

11## References

1. Schmelz, Joan T. (2011, September). *Genesis of the CSWA.* Retrieved from: https://cswa.aas.org/genesis.html

2. National Science Foundation (2018). *National Center for Science and Engineering Statistics, Survey of Earned Doctorates, 2018*. Retrieved from: https://ncses.nsf.gov/pubs/nsf19301/data

3. Ivie, R., & Porter, A.M. (2019, March). *Women in Physics and Astronomy, 2019*. American Institute of Physics. Retrieved from: https://www.aip.org/sites/default/files/statistics/women/Women%20in%20Physics%20and%20Astronomy%202019.1.pdf

4. National Academies of Sciences, Engineering, and Medicine. (2018, June). *Sexual harassment of women: climate, culture, and consequences in academic sciences, engineering, and medicine*. National Academies Press. Retrieved from: https://www.nap.edu/read/24994/chapter/2

5. Ivie R., White S., and Chu R. (2017, July). *Demographics and Intersectionality in Astronomy*. American Institute of Physics. Retrieved from: https://www.aip.org/statistics/reports/demographics-and-intersectionality-astronomy

6. Clancy, K. B., Lee, K. M., Rodgers, E. M., & Richey, C. (2017, July). *Double jeopardy in astronomy and planetary science: Women of color face greater risks of gendered and racial harassment.* Journal of Geophysical Research: Planets. Retrieved from: https://agupubs.onlinelibrary.wiley.com/doi/full/10.1002/2017JE005256

7. American Physical Society. (2016, March). *LGBT Climate in Physics: Building an Inclusive Community.* Retrieved from: https://www.aps.org/programs/lgbt/upload/LGBTClimateinPhysicsReport.pdf

8. Rathbun, J. (2016, September). *Women on Spacecraft Missions: Are we moving towards parity with the percentage in the field?* The Planetary Society. Retrieved from: http://www.planetary.org/blogs/guest-blogs/2016/0916-women-on-spacecraft-missions.html